# Source -Free Domain Adaptation for Speaker Verification in Data-Scarce Languages and Noisy Channels

*Shlomo Salo Elia, Aviad Malachi, Vered Aharonson, Gadi Pinkas*
*Afeka Center for Language Processing*
*and Afeka Tel-Aviv College of Engineering*

**Abstract.** Domain adaptation is often hampered by exceedingly small target datasets and inaccessible source data. These conditions are prevalent in speech verification, where privacy policies and/or languages with scarce speech resources limit the availability of sufficient data. This paper explored techniques of source-free domain adaptation unto a limited target speech dataset for speaker verification in data-scarce languages. Both language and channel mis-match between source and target were investigated. Fine-tuning methods were evaluated and compared across different sizes of labeled target data. A novel iterative cluster-learn algorithm was studied for unlabeled target datasets. All fine-tuning methods outperformed the "from-scratch" trainings and a pre-trained benchmarks in 1 to 100 hours of labelled-target data sizes. A Siamese neural network exhibited slightly better results compared to the speaker-identification fine-tuning methods in the smallest target datasets. The iterative cluster-learn algorithm performance in the 100-hours unlabeled target was as effective as the supervised methods performance on the smaller labeled datasets. These preliminary results imply a potential advantage of a Siamese-network and iterative clustering in very-small labeled datasets of speakers, for supervised and unsupervised domain adaptation, respectively.

## 1 Introduction

Domain adaptation (DA) aims to improve the performance of a domain model by using the knowledge from an available source domain when the source and target domains stem from different feature spaces or distributions [1]. DA is therefore widely-used in AI applications, and particularly in speaker recognition, where recorded datasets portray a mismatch between source and target sets in both language and recording channel [2].

Speaker recognition includes two major applications: speaker identification (SI) and speaker verification (SV). SI identifies a speaker from a group of known speakers. SV tests the hypothesis that the speaker in two recordings is the same person [3].

SV studies followed image and natural language processing applications and implemented a proliferation of DA methods. Both supervised and unsupervised training were explored, where the supervised methods use speaker labels in the target dataset. Former methods sought to make the data or model domain-independent through statistical methods. Factor analysis [4], correlation alignment (CORAL) [5,6], source-normalization for LDA [7] and within-speaker covariance correction [8]. Recent methods adapted the source unto the target domain, using adversarial training methods [9,10,11], autoencoder [12], a model-based version of correlation alignment [13], and multiple fine tuning strategies [14]. These methods employ deep neural networks (DNN) that learn speaker embeddings - low-dimensional, high-level feature vectors - of a speaker's voice [15,16]. Siamese neural network (SNN) [17], were recently used for generating embeddings and act as a classifier on SV tasks [18, 19].

The above-mentioned methods require direct access to source domain labelled speakers' recordings during training. Real-world scenarios, however, often prohibit the access to source samples due to privacy policies or compute resources [20]. Methods that could address this limitation in SV tasks were evaluated in the domain adaptation challenge 2013 in both supervised [21] and unsupervised conditions [22,23,24]. Their fine-tuning of pre trained out-of-domain source model, was recently referred to as source data free domain adaptation (SFDA) [20].

An additional challenge to SFDA is that target datasets particularly in scare-resources languages are often exceedingly small, smaller than the ones employed in the afore-mentioned studies. These conditions prohibit, in example, the prevalently-used probabilistic linear discriminant analysis adaptation through linear interpolation [21-24]. A clustering process were suggested as an alternative to the linear interpolation [22], but the idea was not explored.

Our study extends the former findings in two aspects. The first is an implementation of the more recent DNN embedding-learning methods to study the extent of target-domain size reduction which would not drastically impact SV performance for SFDA. The second is an investigation of the proposed iterative cluster-learn algorithm for unsupervised SFDA in data-scarce target conditions. Our source domain pretraining entailed clean recording of English speakers. The target domain included Levantine Arabic, Farsi, Fastu, Dari and Urdu speakers, whose recordings were retransmitted through two different noisy channels. SFDA for down-sized target datasets, between 1 and 25 hours of speech, was investigated.

## 2 Methods

### 2.1 Data

The VoxCeleb [25, 26] English speakers' dataset was used as a source domain. DARPA's robust automatic transcription of speech (RATS) SI corpus was used for the target datasets generation. This corpus includes conversational telephone speech recordings and annotation data to provide training, development, and initial test sets for SI [27]. The dataset included native speakers of Levantine Arabic, Farsi, Dari, Pashto, and Urdu, is presented in Table 1. The original audio files were retransmitted through eight different noisy communication channels, labeled by the letters A through H. Only the channel A - ultra-high frequency retransmission, and channel D - high frequency retransmission channels were used in our experiments. The original recordings, referred to in the RATS documentation as "source" files, were referred to in our study as "initial", to avoid confusion with the source and target terms of DA. Our study used the RATS SI development subset for the generation of enrollment and testing samples, and its training subset for the training samples.

### 2.2 Preprocessing

Each initial RATS recording underwent a voice activity detection (VAD), that yielded output recordings containing only speech segments. Subsequently, output recordings shorter than 8 seconds were filtered out. The remaining recording files were divided into segments of 8 seconds. Speakers in the training and development subsets with less than 3 and 2 segments respectively, were removed from the dataset. Table 2 describes the properties of the development and training subsets before and after pre-processing, in the initial (I), A and D channels.

### 2.3 Datasets preparation

The training datasets of each channel included 45,000 segments files that amounted to a 100 hours of speech. The segments were randomly chosen from the training subset and the segments per speaker ratio (R/S) was balanced as much as possible. Table 3 portrays the number of speakers, the mean and standard deviation of segments per speaker and the R/S ratio in the training sets of each channel.

The speakers in the development subset were divided into two sets of equal sizes. The first set was used to create enrollment-test pairs for the Siamese validation-set. The second was used to create the pairs for an evaluation test-set, which was shared by all the experiments. The pairs were randomly generated while balancing label distributions (50% target and 50% non-target). The datasets in each channel contained 30,000 pairs. Table 4 portrays the validation and test sets.

**Speaker Identification (SI) Training and Validation**. The 45,000 segments datasets described in Table 3 were used for the SI training. One segment per speaker was chosen, of approximately 12 minutes total, was chosen for the validation. Data subsets of varying sizes were generated by sorting the speakers according to their ID in ascending order and randomly selecting the first 1%, 2%, 3%, 6%, 12%, 25% in the list. Table 5 portrays the properties of those six subsets in each of the three channels. The number of hours in each subset was approximately equal to the percentage of speakers in the subset, such that 1% ~ 1 hour, 2% ~ 2 hours, and so on. Each subset was thus referred to by the number of hours it contained.

**Table 1.** Dataset language distribution.

| Language | Files | Hours |
|---|---|---|

| Levantine Arabic | 1110 | 224.87 |
|---|---|---|
| Farsi | 988 | 201.23 |
| Dari | 1639 | 327.58 |
| Pashto | 2903 | 586.58 |
| Urdu | 2763 | 557.06 |

**SNN training and validation**. The training datasets for the SNN were created by a random generation of 30,000 balanced pairs of segments from each of the six sub-datasets (1, 2, 3, 6, 12,18 and 25) and the 100 hours set, in each channel. A single validation set was generated for all SNN experiments and included different 30,000 randomly-generated pairs selected from the Siamese-validation subset of Table 4. The target and non-target label distributions in all training and validation sets were equally distributed.

The 45,000 segments enabled a generation of a much larger number of pairs. Preliminary observations from trainings performed on 60, 90, 120, 150 and 180 thousand pairs demonstrated similar results but are not included in this paper.

**Table 2.** Training (Train) and development (Dev) sub-sets properties in channels A, D and initial (I) sets, before and after preprocessing.

| Channel | | Files | | Speakers | | Hours | |
|---|---|---|---|---|---|---|---|
| | | before | after | before | after | before | after |
| A | Train | 5888 | 5480 | 5283 | 4378 | 1140 | 118 |
| | Dev | 2765 | 1885 | 289 | 212 | 564 | 37 |
| D | Train | 6303 | 6241 | 5875 | 5769 | 1265 | 382 |
| | Dev | 3054 | 2151 | 313 | 227 | 624 | 124 |
| I | Train | 6333 | 6268 | 5903 | 5882 | 1271 | 717 |
| | Dev | 3060 | 1947 | 313 | 214 | 625 | 223 |

**Unsupervised training-validation**. The same SI training datasets were used for the unsupervised training, where speaker labels were discarded. Speaker labels were generated using our clustering method, detailed in section 2.5. The validation set was generated by choosing one sample from each clustered-speaker, as in the supervised SI validation-set generation.

**Table 3.** The 100-hours training sets for channels A, D, and the Initial recording, including the number of speakers per channel, the mean and standard deviation (std) of the segments per speaker and the recordings per speaker ratio (R/S).

| Channel | Number of speakers | Mean | Std | R/S |
|---|---|---|---|---|
| A | 4378 | 10.27 | 8.04 | 1.25 |
| D | 5769 | 7.8 | 0.91 | 1.07 |
| Initial | 5882 | 7.65 | 0.47 | 1.07 |

## 2.4 Training Methods

All training methods employed a state-of-the-art embedding model based on the time-delay deep neural networks (TDNN) architecture, using the emphasized channel attention propagation and aggregation architecture (ECAPA) [16]. The main architecture contains three components, including feature-learning component, statistical pooling component and speaker-classification component. Once trained, the 512- dimensional activations of the penultimate layer are read out as the speech embedding [15,28].

**From Scratch Training.** "From-Scratch" speaker-identification (SI) training served as a benchmark for the subsequent fine-tuning methods. The ECAPA-TDNN model [16] was trained on the RATS labeled training sets. The model's weights were initialized randomly and its last SoftMax layer was adjusted to the target training set labels for the SI task [16].

Table 4. The properties of the Siamese network (SNN) validation and test sets for channels A, D, and Initial recordings.

| Channel | SNN set | Speakers | Samples | Mean | Std |
|---|---|---|---|---|---|
| A | test | 105 | 8174 | 77 | 63 |
|   | validation | 105 | 7942 | 75 | 56 |
| D | test | 113 | 27,809 | 246 | 111 |
|   | validation | 113 | 27,181 | 240 | 96 |
| Initial | test | 107 | 50,008 | 467 | 131 |
|   | validation | 107 | 49,585 | 463 | 127 |

Table 5. The properties of the downsized subsets of channels A, D, and Initial recordings.

| Channel | Size | Speakers | Segments | Mean | Std |
|---|---|---|---|---|---|
| A | 1 hour | 68 | 701 | 10.3 | 6.56 |
|   | 2 hours | 102 | 1148 | 11.25 | 7.75 |
|   | 3 hours | 136 | 1497 | 11 | 7.95 |
|   | 6 hours | 273 | 2768 | 10.13 | 7.16 |
|   | 12 hours | 547 | 5903 | 10.79 | 8.27 |
|   | 18 hours | 820 | 10,123 | 12.34 | 9.58 |
|   | 25 hours | 1904 | 13,551 | 12.38 | 9.63 |
| D | 1 hour | 90 | 698 | 7.75 | 1.12 |
|   | 2 hours | 135 | 1049 | 7.77 | 1.03 |
|   | 3 hours | 180 | 1395 | 7.75 | 1.1 |
|   | 6 hours | 360 | 2784 | 7.73 | 1.09 |
|   | 12 hours | 721 | 5625 | 7.8 | 0.93 |
|   | 18 hours | 1081 | 8487 | 7.85 | 0.8 |
|   | 25 hours | 1442 | 11,322 | 7.85 | 0.78 |
| Initial | 1 hour | 91 | 685 | 7.63 | 0.48 |
|   | 2 hours | 137 | 1045 | 7.62 | 0.48 |

| | | | | |
|---|---|---|---|---|
| 3 hours | 183 | 1401 | 7.65 | 0.47 |
| 6 hours | 367 | 2819 | 7.68 | 0.46 |
| 12 hours | 735 | 5611 | 7.63 | 0.48 |
| 18 hours | 1102 | 8376 | 7.6 | 0.48 |
| 25 hours | 1470 | 11,204 | 7.62 | 0.48 |

**SI Training fine-tuning.** The speaker-identification (SI) fine-tuning used the same architecture as the FS training. The model was trained on the RATS training sets for the SI task, using the ECAPA pre-trained weights, and its last layer weights were randomly initialized.

**Siamese Training fine-tuning.** The Siamese neural network (SNN) architecture was selected based on performance comparison in our preliminary trials. The architecture consisted of two VoxCeleb-SI pre-trained ECAPA networks and the embedding layers of the two networks were combined according to equation (1):

$$output = \sigma(W(x_1 \odot x_2) + b) \qquad (1)$$

Where $\sigma$ is the sigmoid function, $x_1$ and $x_2$ are the embedding layers of the first and second ECAPA networks, respectively, $W$ is a weight vector and $b$ is a scalar bias. The two embedding layers are multiplied pairwise. The weights ($W$) and the bias ($b$) were randomly initialized. The fine-tuning training on RATS training sets, for a SV task, employed generated segment-pairs, binary labels, and cross-entropy loss.

**Unsupervised iterative clustering-training.** An iterative clustering-training algorithm was designed and used for unlabeled target datasets. The algorithm entails the following date steps for a trained model $\Phi$:

1. For each recorded segment of the target domain, generate a speaker embedding using $\Phi_i$.
2. Cluster the embeddings and assign each segment a label based on its cluster ID.
3. Split the data into a train set and a hypothesized validation set.
4. Train $\Phi_i$ on the target dataset, using the assigned cluster IDs as labels. Evaluate $\varepsilon_i$.
5. Increment i and assign $\Phi_{i+1} = \Phi_i$
6. Repeat steps 1-5 until $\varepsilon_i < \varepsilon_{i+1}$.

Where $\Phi_i$ is the trained model in the ith iteration and $\varepsilon_i$ is the validation error of $\Phi_i$. $\Phi_0$ and $\varepsilon_0$ are the initial pre-trained model and its validation error, respectively
The clustering in step 2 was implemented using both K-Means and the agglomerative hierarchical clustering (AHC) and the two methods were compared. The training in step 4 employed the SI fine-tuning method. Step 3 reflects the assumption that there are no ground-truth labels to create a verified validation dataset.

The input for the clustering algorithm was generated using two techniques. Technique I treats each 8-second sample segment as an independent recording. Technique II uses the original recording labels of the 8-second segments, and averages all segment embeddings of each original recording. Each original recording in this technique was thus represented by a single averaged embedding and the averaged embeddings were clustered by the algorithm. The cluster IDs of each averaged embedding was assigned to all their segments.

## 3   Experiments

The experiments were separately conducted on the initial RATS recordings and the channels A and D retransmitted recordings. The supervised fine-tuning experiments were repeated on data subsets of 1,2,3,6,12,18,25 and 100 hours, for each channel. The unsupervised clustering experiments were performed only on the 100 hours' datasets of each channel (Table 3).

### 3.1   Implementation Details

The training employed the SpeechBrain toolkit [29], using the hyper-parameters defined in the original ECAPA-TDNN architecture [16]. The VAD's pipeline employed SpeechBrain with default parameters. The learning rate in the SI fine-tuning and the "from-scratch" training was set to 0.001. The model weights (Equation 1) in the first phase of the SNN fine-tuning were frozen up to the embedding layer, and the remaining weights were trained with LR=0.01. In the second phase, all Siamese network's weights were trained using LR=0.001. The number of epochs in the SI training was set to 40, and in the SNN training, to 20. Scikit-Learn library [30] was used for the clustering algorithms' implementation. The number of centroids initializations in the K-Means was set to 10, the maximum iterations to 300 and the method to 'k-means++' [30]. In the AHC experiments, the affinity metric and linkage were set to 'cosine' and 'average', respectively [30].

### 3.2   Evaluation and Baselines

Equal Error Rate (EER) was evaluated on all the test-sets. The EER in the SI tests was based on a cosine similarity score between the two enrolment and the test embeddings. The EER computation in the SNN tests used the score of the network's sigmoid output.

The baselines for each channel were the EERs of pre-trained ECAPA evaluated on the RATS tests sets, without fine-tuning. This model's benchmark performance demonstrated an EER of 0.87% on the VoxCeleb test set [16].

## 4   Results

Tables 6 portrays the EER results of the "from scratch" and the SI fine-tuning in the three RATS channels: Initial, A, and D. Table 7 similarly portrays the EER of the SI and SNN fine-tuning methods. The data subset sizes are denoted by their length in hours: 1,2,3,6,12,18,25 and 100.

Table 8 and 9 depict the EER results in the iterative clustering-training algorithm experiments, for techniques I and II respectively. The numbers 1,2 and 3 indicate the number of iterations.

In all four tables, a bold font indicates the best test EER. The underlined numbers in Tables 8 and 9 indicate the best validation EER results.

**Table 6.** SI EER (%) in the from-scratch (FS) and fine-tuning (FT) experiments, for channels A, D, and the Initial recordings.

| Subset | Initial recordings | | D channel | | A channel | |
|---|---|---|---|---|---|---|
| Baseline | 12.91 | | 14.89 | | 19.96 | |
| - | FS | FT | FS | FT | FS | FT |
| 1 hour | 19.81 | **12.72** | 19.2 | **13.21** | 20.55 | **16.19** |
| 2 hours | 18.94 | **11.47** | 20.83 | **13.09** | 19.32 | **15.96** |
| 3 hours | 16.76 | **10.69** | 22.2 | **12.93** | 19.29 | **14.61** |
| 6 hours | 16.75 | **8.57** | 19.64 | 17.08 | 18.57 | **13.27** |
| 12 hours | 12.67 | **8.3** | 17.15 | **13.56** | 15.67 | **11.47** |
| 18 hours | 11.98 | **8.36** | 15.28 | **12.11** | 13.67 | **10.4** |
| 25 hours | 10.62 | **7.27** | 13.84 | **10.8** | 12.32 | **9.31** |
| 100 hours | 6.94 | **5.61** | 10.61 | **9.06** | 8.23 | **7.41** |

**Table 7.** EER (%) in the SI vs SNN fine-tuning experiments, for channels A, D, and the Initial recordings.

| Subset | Initial recordings | | D channel | | A channel | |
|---|---|---|---|---|---|---|
| baseline | 12.91 | | 14.89 | | 19.96 | |
| - | SI | SNN | SI | SNN | SI | SNN |
| 1 hour | 12.72 | **12.65** | **13.21** | 14.49 | **16.19** | 17.80 |
| 2 hours | 11.47 | **10.96** | **13.09** | 13.43 | **15.96** | 16.54 |
| 3 hours | 10.69 | **10.35** | 12.93 | **12.66** | **14.61** | 16.07 |
| 6 hours | **8.57** | 9.69 | 17.08 | 15.70 | **13.27** | 14.24 |
| 12 hours | **8.30** | 9.67 | **13.56** | 15.41 | **11.47** | 13.76 |
| 18 hours | **8.36** | 9.47 | **12.11** | 13.06 | **10.40** | 11.66 |
| 25 hours | **7.27** | 8.44 | 10.79 | 12.46 | **9.31** | 11.2 |
| 100 hours | **5.61** | 7.87 | **9.06** | 11.73 | **7.41** | 10.28 |

**Table 8.** EER (%) of Technique I in 3 iterations, using K-Means and AHC, for channels A, D, and the Initial recordings.

| Clustering | Channel | Iteration | | |
|---|---|---|---|---|
| | | 1 | 2 | 3 |
| K-MEANS | Initial | 8.73 | <u>**8.21**</u> | 8.26 |
| | A | 13.41 | <u>**10.12**</u> | 11.3 |
| | D | 14.31 | 14.19 | <u>**14.01**</u> |
| AHC | Initial | 9.73 | **8.57** | <u>8.68</u> |

| | A | 14.71 | 14.55 | **14.15** |
|---|---|---|---|---|
| | D | **16.52** | 17.13 | - |

**Table 9.** EER (%) of Technique II in 2 iterations, using K-Means and AHC, for channels A, D, and the Initial recordings.

| Clustering | Channel | Iteration | |
|---|---|---|---|
| | | 1 | 2 |
| K-MEANS | Initial | **5.86** | 6.04 |
| | A | **8.11** | 8.4 |
| | D | **9.98** | 10.97 |
| AHC | Initial | **5.59** | 6.11 |
| | A | **8.43** | 9.03 |
| | D | 10.04 | 11.15 |

All fine-tuning experiments in Table 6 exhibited a smaller EER compared to the "from scratch" training, as well as the ECAPA baseline. A single exception is the 6 hours subset, where the fine tuning EER is higher than the baseline and higher than the EER of all fine-tuned subsets, in all three channels. The differences between the EER of the two fine-tuning methods (Table 7) were all smaller than the differences portrayed in Table 6. The EER of the SNN fine-tuning in the smallest subsets: 1, 2 and 3 hours (Table 7) was slightly smaller than both baseline and SI fine-tuning.

All EER in the unsupervised experiments using technique II exceed the baseline (bolded values, Table 9) already in the first iteration. The EER in technique I exceed the baseline in one of the three iterations (Table 8).

## 5 Discussion

Two aspects of domain adaptation for SV were explored, for the conditions where only an out-of-domain pre-trained source model and small target dataset are available. The first entailed supervised speaker verification for incrementally down-sized target data. The results indicated that fine tuning outperformed the "from-scratch" training in all experiments (Table 6). Moreover, all training methods outperformed a benchmark network, which previously performed exceedingly well on the source domain. This trend persisted in both initial clean recording data and retransmitted noisy data. These preliminary findings imply the potential of our methods in the challenging conditions of central- and southern-Asian languages, noisy channels and very-small datasets. Another indication in support of this hypothesis is that the differences between the different training-methods performance decreased as their training size increased.

An exception to these trends were the experiments on of the 6-hours channel D recordings, where neither SI nor SNN fine tuning improved the baseline, and both performance were worse than for the smaller 1, 2 and 3 hours' experiments. The SNN addi-tionally exhibited the same out-of-trend performance in the 12 hours' channel D experiments. Both results may stem from a sampling mismatch between training and test in this sub-set. This assumption should be investigated in future studies.

The preliminary results in Table 7 imply a potential of the SNN to outperform SI fine-tuning for small target datasets that contain a very small number of speakers. This result was observed in the source channel 1-3 hours and the D channel 3-hours experiments. The SI fine-tuning, however, yielded better results in all channel A experiments. No definite conclusions could therefore be drawn on the effectiveness of the SNN's in these tasks, and the results may stem from the arbitrary sampling of the small datasets. More experiments and enhancements such as "triplets" architecture may further improve the SNN-fine-tuning performance.

The unsupervised SV experiments yielded that the iterative clustering-training algorithm outperformed the baselines. Technique II (Table 9) portrayed a performance similar to the fine-tuning within one iteration. Technique I outperformed the baseline, but to a lesser degree. Technique I, however, may reflect a larger recordings/speaker ratio (R/S) in the dataset. This ratio was an order of magnitude larger in technique I: R/S was 7.65 and 1.07 in the Source channel for Technique I and II respectively, 7.8, and 1.07 in channel D and 10.27 and 1.25 in channel A. Although technique II outperformed technique I and required less computation resources, the results of table 8 may better reflect target datasets where the recordings/speaker ratio is greater than in the original RATS dataset. In addition, using a reliably-labeled validation set instead of the one based on clustering, is likely to improve the algorithm's stop criteria.

The present study assumed that K, the number of speakers, is known. In the future, we aim to determine K by evaluating the embeddings' quality of different K's for the first iteration.

The study implied a potential of Siamese fine-tuning in supervised DA on very-small labeled datasets. Unsupervised iterative clustering was found as effective as supervised DA.

## References

1. Farahani, A., et al.: A brief review of domain adaptation. In: Advances in Data Science and Information Engineering: Proceedings from ICDATA 2020 and IKE 2020, pp. 877-894 (2021).
2. Bahmaninezhad, F., et al.: An investigation of domain adaptation in speaker embedding space for speaker recognition. Speech Communication 129, 7-16 (2021).
3. Reynolds, D.A.: An overview of automatic speaker recognition technology. In: Acoustics, speech, and signal processing (ICASSP), IEEE international conference on. IEEE, vol. 4, pp. IV–4072 (2002).
4. Dehak, N., Kenny, P.J, Dehak, R., Dumouchel, P., Ouellet, P.: Front-end factor analysis for speaker verification. In: IEEE Transactions on Audio, Speech, and Language Processing, vol. 19, no. 4, pp. 788–798 (2010).
5. Sun, B., Feng, J., Saenko, K.: Return of frustratingly easy domain adaptation. In: Proceedings of the AAAI Conference on Artificial Intelligence, vol. 30, no. 1 (2016).
6. Alam, J., Bhattacharya, G., Kenny, P.: Speaker verification in mismatched conditions with frustratingly easy domain adaptation. In: Proceedings Odyssey, pp. 176 – 180 (2018).

7. M. McLaren, D., Van Leeuwen, D.: Source-normalized LDA for robust speaker recognition using i-vectors from multiple speech sources. In: IEEE Transactions on Audio, Speech, and Language Processing, vol. 20, no. 3, pp. 755–766 (2011).
8. Glembek, O., Ma, J., Matejka, P., Zhang, B., Plchot, O., Burget, L., Matsoukas, S.: Domain adaptation via within-class covariance correction in i-vector based speaker recognition systems. In: Proceedings IEEE International Conference on Acoustics, Speech and Signal Processing, pp. 4032–4036 (2014).
9. Wang, Q., Rao, W., Sun, S., Xie, L., Chng, E.S, Li, H.: Unsupervised domain adaptation via domain adversarial training for speaker recognition. In: IEEE International Conference on Acoustics, Speech and Signal Processing (ICASSP 2018), pp. 4889–4893 (2018).
10. Xia, W., Huang, J., Hansen, J.H.L.: Cross-lingual text-independent speaker verification using unsupervised adversarial discriminative domain adaptation. In: IEEE International Conference on Acoustics, Speech and Signal Processing, pp. 5816–5820 (2019).
11. Chen, Z., Wang, S., Qian, Y.: Adversarial domain adaptation for speaker verification using partially shared network. In: Proceedings ISCA InterSpeech, pp. 3017–3021 (2020).
12. Wang, X., Li, L., Wang, D.: VAE-based domain adaptation for speaker verification. In: 2019 Asia-Pacific Signal and Information Processing Association Annual Summit and Conference (APSIPA ASC), pp. 535-539 (2019).
13. Lee, K.A, Wang, Q., Koshi-naka, T.: The CORAL+ algorithm for unsupervised domain adaptation of PLDA: In: IEEE International Conference on Acoustics, Speech and Signal Processing (ICASSP). IEEE, pp. 5821–5825 (2019).
14. Zhang, C., et al.: An Analysis of Transfer Learning for Domain Mismatched Text-independent Speaker Verification. In: Proceedings Odyssey Workshop, pp. 181–186 (2018).
15. Snyder, D., Garcia-Romero, D., Sell, G., Povey, D., Khudanpur, S.: X-vectors: Robust DNN embeddings for speaker recognition. In: IEEE International Conference on Acoustics, Speech and Signal Processing, pp. 5329–5333 (2018).
16. Desplanques, B., Thienpondt, J., Demuynck, K.: ECAPA-TDNN: emphasized channel attention, propagation and aggregation in TDNN based speaker verification. In: Proceedings of Interspeech, pp. 3830–3834 (2020).
17. Chicco, D.: Siamese neural networks: An overview. Artificial Neural Networks. Humana, New York (2020).
18. Zhang, Y., Yu, M., Li, N., Yu, C., Cui, J., Yu, D.: Seq2seq attentional siamese neural networks for text-dependent speaker verification. In: IEEE International Conference on Acoustics, Speech and Signal Processing, pp. 6131–6135 (2019).
19. Balipa, M., Farhath, A.: Twins Voice Verification and Speaker Identification. In: 2022 International Conference on Artificial Intelligence and Data Engineering (AIDE), pp. 322-325. IEEE (2022).
20. Kim, Y., et al.: Domain adaptation without source data. IEEE Transactions on Artificial Intelligence **2**(6), 508-518 (2021).
21. Garcia-Romero, D., Mccree, A.: Supervised domain adaptation for i-vector based speaker recognition. In: IEEE International Conference on Acoustics, Speech and Signal Processing, pp. 4047–4051 (2014).
22. Shum, S., Reynolds, D.A, Garcia-Romero, D., McCree, A.: Unsupervised Clustering Approaches for Domain Adaptation in Speaker Recognition Systems. In: Proceedings of Odyssey 2014 - The Speaker and Language Recognition Workshop, pp. 265–272 (2014).
23. Garcia-Romero, D., McCree, A., Shum, S., Brummer, N., Vaquero, C.: Unsupervised Domain Adaptation for I-Vector Speaker Recognition. In: Proceedings of Odyssey 2014 - The Speaker and Language Recognition Workshop, pp. 260– 264 (2014).

24. Garcia-Romero, D., Zhang, X., McCree, A., Povey, D.: Improving speaker recognition performance in the domain adaptation challenge using deep neural networks. In: Spoken Language Technology Workshop (SLT) IEEE. IEEE, pp. 378–383 (2014).
25. Nagrani, A., Chung, J.S, Zisserman, A.: VoxCeleb: A largescale speaker identification dataset. In: Proceedings Interspeech, pp. 2616–2620 (2017).
26. Chung, J.S, Nagrani, A., Zisserman, A.: VoxCeleb2: Deep speaker recognition. In: Proceedings Interspeech, pp. 1086–1090 (2018).
27. Graff, D., Ma, X., Strassel, S. , Walker, K., Jones, K. RATS Speaker Identification LDC2021S08. Web Download. Philadelphia: Linguistic Data Consortium (2021).
28. Okabe, K., Koshinaka, T., Koichi S.: Attentive statistics pooling for deep speaker embedding. In: Intrespeech 2018, pp. 2252–2256 (2018).
29. Ravanelli, M., et al.: SpeechBrain: A General-Purpose Speech Toolkit. arXiv:2106.04624, (2021).
30. Pedregosa, F., et al.: Scikit-learn: Machine Learning in Python. Journal of Machine Learning Research 12, 2825-2830 (2011).